\magnification=1200

\noindent

\null

\centerline{\bf ON THE DAMPING OF THE ANGULAR MOMENTUM}
\centerline{\bf OF THREE HARMONIC OSCILLATORS}

\vskip 1truecm

\centerline{A. Isar, A. Sandulescu}
\centerline{Department of Theoretical Physics, Institute
of Physics and Nuclear Engineering}
\centerline{Bucharest-Magurele, Romania}

\vskip 1truecm

\centerline{ABSTRACT}

In the frame of the Lindblad theory of open quantum systems, the system of
three uncoupled harmonic oscillators with opening operators linear in the
coordinates and momenta of the considered system is analyzed. The damping
of the angular momentum and of its projection is obtained.

\vskip 0.5truecm

{\bf 1. Introduction}

In the last two decades, more and more interest arose about the problem of
dissipation in quantum mechanics, i.e. the consistent description of open
quantum systems [1-6].

Because dissipative processes imply irreversibility, and, therefore, a
preferred direction in time, it is generally thought that quantum dynamical
semigroups are the basic tools to introduce dissipation in quantum mechanics.
The most general form of the generators of such semigroups (under some
topological conditions) was given by Lindblad [7-9], whose formalism has
recently been applied to various physical phenomena, for instance, to the
damping of collective modes in deep inelastic collisions [10-15]. An important
feature of these reactions is the dissipation of energy and angular momentum
out of the collective degrees of freedom into the intrinsic or single-particle
degrees of freedom.

The damping of energy and angular momentum can be described at different levels
of approximations: one may use quantum or classical mechanical methods and
microscopical or phenomenological models.

In the previous papers [13-15] simple phenomenological models for the damping
of the angular momentum were studied in the framework of the Lindblad quantum
mechanical theory. In the present paper, for a special choice of the opening
coefficients we obtain explicit expressions for the damping of the angular
momentum and projection of the angular momentum of three harmonic oscillators.
As opening operators we use the generators of the Heisenberg group, the
coordinates $q_k$ and momenta $p_k (k=1,2,3)$. In order to keep the symmetry of
the system we assume that the opening coefficients are symmetric under the
permutation of the axes. In this way we include the influence of the coupling,
due to the environment, of the three harmonic oscillators, initially uncoupled.

The paper is organized as follows: In Sec.2 we first present the Lindblad
equation of motion in the Heisenberg picture. In Sec.3 we obtain the damping
of the projection of the angular momentum for a system of three uncoupled
harmonic oscillators. In Sec.4 we analyze the behaviour of the squared angular
momentum for the same system.

\vskip 0.5truecm

{\bf 2. The Lindblad formalism in the Heisenberg picture}

In Lindblad's formalism, the usual von Neumann-Liouville equations ruling the
time evolution of closed quantum systems are replaced by the following ones
[7-9]:
$${d\Phi_t(\rho)\over dt}=L[\Phi_t(\rho)],$$
$${d\widetilde\Phi_t(A)\over dt}=\widetilde L[\widetilde\Phi_t(A)],\eqno(2.1)$$
in the Schr\"odinger and Heisenberg picture, respectively. Here,
$\Phi_t(\widetilde\Phi_t)$ is the dynamical semigroup describing the
irreversible time evolution of the open system in the Schr\"odinger
(Heisenberg)
representation; $\rho$ is the density operator and $A$ any operator acting on
the Hilbert space ${\bf H}(A\in{\bf B}({\bf H}))$; finally, $L(\widetilde L)$
is the
infinitesimal generator of the dynamical semigroup $\Phi_t(\widetilde\Phi_t)$.

By using the Lindblad theorem [7-9] which gives the most general form taken
by the generator $\widetilde L$, the Markovian master equation (2.1) takes the
form:
$${d\widetilde\Phi_t(A)\over dt}=\widetilde L(\widetilde\Phi_t(A))={i\over
\hbar}[H,\widetilde\Phi_t(A)]+{1\over 2\hbar}\sum_j(V_j^+[\widetilde\Phi_t(A),
V_j]+[V_j^+,\widetilde\Phi_t(A)]V_j).\eqno(2.2)$$
Here, $H$ is the Hamiltonian of the system and the operators $V_j,V_j^+$ are
taken as polynomials of only first degree in the observables $q_1,q_2,q_3,p_1,
p_2,p_3$, which represent the hermitian generators of the Heisenberg group.
Then in the linear space spanned by $q_k,p_k (k=1,2,3)$, there exist only six
linearly independent operators $V_{j=1,2,...,6}$:
$$V_j=\sum_{k=1}^3 a_{jk}p_k+\sum_{k=1}^3 b_{jk}q_k,$$
where $a_{jk},b_{jk}\in {\bf C}$ with $j=1,2,...,6$.

Then it yields
$$V_j^+=\sum_{k=1}^3 a_{jk}^*p_k+\sum_{k=1}^3 b_{jk}^*q_k,$$
where $a_{jk}^*,b_{jk}^*$ are the complex conjugates of $a_{jk},b_{jk}$.

The coordinates $q_k$ and the momenta $p_k$ obey the usual commutation
relations $(k,l=1,2,3)$:
$$[q_k,p_l]=i\hbar\delta_{kl},~~[q_k,q_l]=[p_k,p_l]=0.$$

Inserting the operators $V_j$ and $V_j^+$ into (2.2) we obtain:
$$\widetilde L(A)={i\over\hbar}[H,A]+{1\over 2\hbar}\sum_{k,l}\{({2\over\hbar}
D_{q_kq_l}-i\alpha_{kl})(p_k[A,p_l]-[A,p_k]p_l)+$$
$$+({2\over\hbar}D_{p_kp_l}-i\beta_{kl})(q_k[A,q_l]-[A,q_k]q_l)-{2\over\hbar}
D_{q_kp_l}(p_k[A,q_l]-[A,p_k]q_l+q_l[A,p_k]-[A,q_l]p_k)-$$
$$-i\lambda_{kl}(p_k[A,q_l]-[A,p_k]q_l-q_l[A,p_k]+[A,q_l]p_k)\}.$$

Here we used the following abbreviations:
$$D_{q_kq_l}=D_{q_lq_k}={\hbar\over 2}Re({\bf a}_k^*\cdot{\bf a}_l),$$
$$D_{p_kp_l}=D_{p_lp_k}={\hbar\over 2}Re({\bf b}_k^*\cdot{\bf b}_l),$$
$$D_{q_kp_l}=D_{p_lq_k}=-{\hbar\over 2}Re({\bf a}_k^*\cdot{\bf b}_l),\eqno(2.3)
$$
$$\alpha_{kl}=-\alpha_{lk}=-Im({\bf a}_k^*\cdot{\bf a}_l),$$
$$\beta_{kl}=-\beta_{lk}=-Im({\bf b}_k^*\cdot{\bf b}_l),$$
$$\lambda_{kl}=-Im({\bf a}_k^*\cdot{\bf b}_l).$$

The scalar products are formed with the vectors ${\bf a}_k,{\bf b}_k$ and their
complex conjugates ${\bf a}_k^*,{\bf b}_k^*$. The vectors have the components
$${\bf a}_k=(a_{1k},a_{2k},...,a_{6k}),$$
$${\bf b}_k=(b_{1k},b_{2k},...,b_{6k}).$$

\vskip 0.5truecm

{\bf 3. The projection of the angular momentum}

The general Hamiltonian of three uncoupled oscillators is
$$H=\sum_{k=1}^3({1\over 2m_k}p_k^2+{m_k\omega_k^2\over 2}q_k^2).$$

The time-dependent expectation values of self-adjoint
operators $A$ and $B$ can be written with the density
operator $\rho$, describing the initial state of the
quantum system, as follows:
$$m_A(t)={\rm Tr}(\rho\widetilde\Phi_t(A)),$$
and, respectively,
$$\sigma_{AB}(t)={1\over 2}{\rm Tr}(\rho\widetilde\Phi_t(AB+BA)).$$

In the following we denote the vector with the six
components $m_{q_i}(t), m_{p_i}(t), i=1,2,3$, by ${\bf
m}(t)$ and the following $6\times 6$ matrix by
$\hat\sigma(t)(i,j=1,2,3)$:
$$\hat\sigma(t)=\left(\matrix{\sigma_{q_iq_j}&\sigma_{q_ip_j}\cr
                            \sigma_{p_iq_j}&\sigma_{p_ip_j}\cr}\right).$$

Then via direct calculation of $\widetilde L(q_k)$ and $\widetilde L(p_k)$ we
obtain
$${d{\bf m}\over dt}=\hat Y{\bf m},\eqno(3.1)$$
where
$$\hat Y=\left(\matrix{-\lambda_{11}&-\lambda_{12}&-\lambda_{13}&{1\over
m_1}&-\alpha_{12}&-\alpha_{13}\cr
-\lambda_{21}&-\lambda_{22}&-\lambda_{23}&-\alpha_{21}&{1\over m_2}&-
\alpha_{23}\cr
-\lambda_{31}&-\lambda_{32}&-\lambda_{33}&-\alpha_{31}&-\alpha_{32}&{1\over
m_3}\cr
-m_1\omega_1^2&\beta_{12}&\beta_{13}&-\lambda_{11}&-\lambda_{21}&-\lambda_{31}
\cr
\beta_{21}&-m_2\omega_2^2&\beta_{23}&-\lambda_{12}&-\lambda_{22}&-\lambda_{32}
\cr
\beta_{31}&\beta_{32}&-m_3\omega_3^2&-\lambda_{13}&-\lambda_{23}&-\lambda_{33}
\cr}\right).\eqno(3.2)$$
From (3.1) it follows that
$${\bf m}(t)=\hat M(t){\bf m}(0)=\exp(t\hat Y){\bf m}(0),\eqno(3.3)$$
where ${\bf m}(0)$ is given by the initial conditions. The matrix $\hat
M(t)$ has to fulfil the condition
$$\lim_{t\to\infty}\hat M(t)=0.\eqno(3.4)$$
In order that this limit exists, $\hat Y$ must have only eigenvalues with
negative real parts.

By direct calculation of $\widetilde L(q_kq_l),\widetilde L(p_kp_l)$ and
$\widetilde L(q_kp_l+p_lq_k),(k,l=1,2,3),$ we obtain
$${d\hat \sigma\over dt}=\hat Y\hat \sigma+\hat \sigma\hat Y^T+2\hat D,
\eqno(3.5)$$
where $\hat D$ is the matrix of the diffusion coefficients $(i,j=1,2,3)$
$$\hat D=\left(\matrix{D_{q_iq_j}&D_{q_ip_j}\cr
D_{p_iq_j}&D_{p_ip_j}\cr}\right)$$
and $\hat Y^T$ the transposed matrix of $\hat Y$. The time-dependent
solution of
(3.5) can be written as
$$\hat \sigma(t)=\hat M(t)(\hat \sigma(0)-\hat \Sigma)\hat M^T(t)+\hat
\Sigma,\eqno(3.6)$$ where $\hat M(t)$ is defined in
(3.3). The matrix $\hat\Sigma$ is time independent and
solves the static problem (3.5) $(d\hat\sigma/dt=0)$:
$$\hat Y\hat\Sigma+\hat\Sigma\hat Y^T+2\hat D=0.\eqno(3.7)$$
Now we assume that the following limit exists for $t\to\infty$:
$$\hat \sigma(\infty)=\lim_{t\to\infty}\hat\sigma(t).\eqno(3.8)$$
In that case it follows from (3.6) and (3.4):
$$\hat\sigma(\infty)=\hat\Sigma.\eqno(3.9)$$
Inserting (3.9) into (3.6) we obtain the basic equations for our purposes:
$$\hat\sigma(t)=\hat M(t)(\hat\sigma(0)-\hat\sigma(\infty))\hat M^T(t)+
\hat\sigma(\infty),
\eqno(3.10)$$
where
$$\hat Y\hat\sigma(\infty)+\hat\sigma(\infty)\hat Y^T=-2\hat D.\eqno(3.11)$$

For the calculation of the matrix $\hat M(t)$ we must diagonalize the matrix
$\hat Y$ by solving the corresponding secular equation, i.e. $\det(\hat Y
-z\hat I)=0$,
where $z$ is the eigenvalue and $\hat I$ is the unit matrix. According to (3.2)
one obtains an equation of sixth order for the eigenvalues $z$, which can be
simply solved only for special examples. In the particular case with
$\alpha_{kl}=0,\beta_{kl}=0,\lambda_{kl}=0(k\not=l)$, the secular equation is
obtained as
$$[(z+\lambda_{11})^2+\omega_1^2][(z+\lambda_{22})^2+\omega_2^2][(z+\lambda_
{33})^2+\omega_3^2]=0.$$
The roots of this equation are
$$z_{1,4}=-\lambda_{11}\pm i\omega_1, z_{2,5}=-\lambda_{22}\pm i\omega_2,
z_{3,6}=-\lambda_{33}\pm i\omega_3.\eqno(3.12)$$
Only positive values of $\lambda_{11},\lambda_{22},\lambda_{33}$ fulfil (3.4).
Applying the eigenvalues $z_i$ of $\hat Y$ we can write the time-dependent
matrix
$\hat M(t)$ as follows:
$$M_{mn}(t)=\sum_i N_{mi}\exp(z_it)N_{in}^{-1},$$
where the matrix $\hat N$ represents the eigenvectors of $\hat Y$:
$$\sum_n Y_{mn}N_{ni}=z_iN_{mi}.$$
With the relations $M_{mn}(t=0)=\delta_{mn}$ and $dM_{mn}(t)/dt\vert_{t=0}=
Y_{mn}$ and using (3.3),(3.10) we conclude that the expectation values of the
coordinates and momenta decay with the exponential factors $\exp(-\lambda_
{11}t)$,$\exp(-\lambda_{22}t)$ and $\exp(-\lambda_{33}t)$ and the matrix
elements $\sigma_{mn}$ with the combined factors $\exp(-2\lambda_{11}t)$,
$\exp(-\lambda_{22}t)$,$\exp(-\lambda_{33}t)$,$\exp[-(\lambda_{11}+
\lambda_{22})t]$, $\exp[-(\lambda_{11}+\lambda_{33})t]$ and $\exp[-(\lambda_
{22}+\lambda_{33})t].$

We present here the matrix $\hat M(t)$ only for our special and simple case
that
the oscillators are uncoupled. With the roots given in (3.12) we obtain
$$\hat M(t)=\left(\matrix{M_{1}&0&0&M_{11}&0&0\cr
0&M_{2}&0&0&M_{12}&0\cr
0&0&M_{3}&0&0&M_{13}\cr
M_{21}&0&0&M_{1}&0&0\cr
0&M_{22}&0&0&M_{2}&0\cr
0&0&M_{23}&0&0&M_{3}\cr}\right),\eqno(3.13)$$
where $(k=1,2,3)$
$$M_{k}=\exp(-\lambda_{kk}t)\cos\omega_kt,$$
$$M_{1k}={1\over m_k\omega_k}\exp(-\lambda_{kk}t)\sin\omega_kt,\eqno(3.14)$$
$$M_{2k}=-m_k\omega_k\exp(-\lambda_{kk}t)\sin\omega_kt.$$
This matrix can be used to evaluate $\hat\sigma(t)$
defined by (3.10). Since we are interested to obtain the
expectation value of the projection of the angular
momentum, we write down the following expressions for
$\sigma_{ q_1p_2}$ and $\sigma_{q_2p_1}$ with $\hat M(t)$
of (3.13),(3.14):
$$\sigma_{q_1p_2}(t)=\exp[-(\lambda_{11}+\lambda_{22})t]((\sigma_{q_1p_2}(0)-
\sigma_{q_1p_2}(\infty))\cos\omega_1t\cos\omega_2t+$$
$$+{1\over m_1\omega_1}(\sigma_{p_1p_2}
(0)-\sigma_{p_1p_2}(\infty))\sin\omega_1t\cos\omega_2t-m_2\omega_2(\sigma
_{q_1q_2}(0)-\sigma_{q_1q_2}(\infty))\cos\omega_1t\sin\omega_2t-$$
$$-{m_2\omega_2\over
m_1\omega_1}(\sigma_{q_2p_1}(0)-\sigma_{q_2p_1}(\infty))\sin\omega_1t\sin\omega
_2
t)+\sigma_{q_1p_2}(\infty),\eqno(3.15)$$
$$\sigma_{q_2p_1}(t)=\exp[-(\lambda_{11}+\lambda_{22})t]((\sigma_{q_2p_1}(0)-
\sigma_{q_2p_1}(\infty))\cos\omega_1t\cos\omega_2t-$$
$$-m_1\omega_1(\sigma_{q_1q_2}
(0)-\sigma_{q_1q_2}(\infty))\sin\omega_1t\cos\omega_2t+{1\over
m_2\omega_2}
(\sigma_{p_1p_2}(0)-\sigma_{p_1p_2}(\infty))\cos\omega_1t\sin\omega_2t-$$
$$-{m_1\omega_1\over
m_2\omega_2}(\sigma_{q_1p_2}(0)-\sigma_{q_1p_2}(\infty))\sin\omega_1t\sin\omega
_2
t)+\sigma_{q_2p_1}(\infty).\eqno(3.16)$$
Similar expressions are found for the other matrix elements of $\hat\sigma(t)$.
The matrix elements of $\hat\sigma(\infty)$ depend on $\hat Y$ and $\hat D$
and must be
evaluated with (3.11) or by the relation [11]:
$$\hat\sigma(\infty)=2\int_0^\infty \hat M(t')\hat D\hat M^T(t')dt'.$$
We obtain that $$\sigma_{q_1p_2}(\infty)=\sigma_{q_2p_1}(\infty).\eqno(3.17)$$

The expectation value of the projection of the angular momentum can be written
from (3.15) and (3.16):
$$<L_3(t)>=\sigma_{q_1p_2}(t)-\sigma_{q_2p_1}(t).\eqno(3.18)$$ If
the three uncoupled oscillators have the same mass and
frequency, then we obtain
$$<L_3(t)>=(\sigma_{q_1p_2}(0)-\sigma_{q_2p_1}(0))\exp[-(\lambda_{11}+\lambda
_{22})t].\eqno(3.19)$$
If we consider the open system to be symmetric, then from (2.3)we have
$\lambda_{11}=\lambda_{22}=\lambda$ and (3.19) becomes:
$$<L_3(t)>=<L_3(0)>\exp(-2\lambda t),$$
with $<L_3(t)>\vert_{t\to\infty}\to 0$ if $\lambda>0$.
This evolution law for the projection of the angular momentum is identical
to that obtained in [14] by another method. We should like to mention that the
same evolution law for the projection $<L_3(t)>$ of the angular momentum can
also be obtained for only two uncoupled harmonic oscillators.

\vskip 0.5truecm

{\bf 4. The squared angular momentum}

In this Section we should like to discuss the behaviour of the squared angular
momentum $<L^2(t)>$ for a system of three independent damped harmonic
oscillators. In this case the expectation value of the angular momentum can be
obtained directly from the following expression:
$$<L^2(t)>=<p_1^2(t)><q_2^2(t)>+<p_1^2(t)><q_3^2(t)>+<p_2^2(t)><q_1^2(t)>+$$
$$+<p_2^2(t)><q_3^2(t)>+<p_3^2(t)><q_1^2(t)>+<p_3^2(t)><q_2^2(t)>-$$
$$-{1\over 2}<(p_1q_1+q_1p_1)(t)><(p_2q_2+q_2p_2)(t)>-{1\over 2}<(p_2q_2+
q_2p_2)(t)><(p_3q_3+q_3p_3)(t)>-$$
$$-{1\over 2}<(p_3q_3+q_3p_3)(t)><(p_1q_1+q_1p_1)(t)>-{3\over 2}\hbar^2.
\eqno(4.1)$$
The expectation values in (4.1) can be obtained by using the expressions for
the centroids and variances for the one-dimensional harmonic oscillator
obtained in [10]. We have $(k=1,2,3)$:
$$<p_k^2(t)>=\sigma_{(pp)_k}(t)+\sigma_{p_k}^2(t),$$
$$<q_k^2(t)>=\sigma_{(qq)_k}(t)+\sigma_{q_k}^2(t),\eqno(4.2)$$
$${1\over 2}<(p_kq_k+q_kp_k)(t)>=\sigma_{(pq)_k}(t)+\sigma_{p_k}(t)\sigma_{q_k}
(t).$$

For each harmonic oscillator the Hamiltonian is chosen of the form (k=1,2,3)
$$H_k={1\over 2m_k}p_k^2+{m_k\omega_k^2\over 2}q_k^2+{\mu_k\over 2}(p_kq_k+
q_kp_k)$$
and the operators in the Lindblad generator can be written in the form
$$V_i^{(k)}=a_i^{(k)}p_k+b_i^{(k)}q_k,i=1,2,$$
with $a_i^{(k)},b_i^{(k)}$ complex numbers.

In the following we consider, for simplicity, the system of three harmonic
oscillators with the same mass $m$, the same frequency $\omega$ and the same
opening coefficients $D_{qq}, D_{pp}, D_{pq}$ and $\lambda$. In addition, we
take $\mu=0$. Then, by introducing the expressions (4.2) taken from [10] into
(4.1), we obtain, after long, but straightforward calculations, the following
form for the expectation value of the squared angular momentum:
$$<L^2(t)>=\beta e^{-4\lambda t}+$$
$$+e^{-2\lambda t}\{\beta_1\sin^2\omega t+\beta_2\cos^2\omega t+\beta_3
\sin\omega t+\beta_4\cos\omega t+\beta_5\sin\omega t\cos\omega t\}+$$
$$+L^2(\infty),$$
where $\beta_j(j=1,...,5)$ are constants. The asymptotic value (if $\lambda>0$)
is given by
$$<L^2(\infty)>=\sigma_{(pp)_1}(\infty)\sigma_{(qq)_2}(\infty)+\sigma_{(qq)_1}
(\infty)
\sigma_{(pp)_2}(\infty)-2\sigma_{(pq)_1}(\infty)\sigma_{(pq)_2}(\infty)+
(cyclic)-$$
$$-{3\over 2}\hbar^2=6(\sigma_{pp}(\infty)\sigma_{qq}(\infty)-
\sigma_{pq}^2(\infty))-{3\over 2}\hbar^2,$$
where [10]
$$\sigma_{qq}(\infty)={1\over 2(m\omega)^2\lambda(\lambda^2+\omega^2)}
((m\omega)^2(2\lambda^2+\omega^2)D_{qq}+\omega^2D_{pp}+2m\omega^2\lambda
D_{pq}),$$
$$\sigma_{pp}(\infty)={1\over 2\lambda(\lambda^2+\omega^2)}((m\omega)^2\omega^2
D_{qq}+(2\lambda^2+\omega^2)D_{pp}-2m\omega^2\lambda D_{pq}),$$
$$\sigma_{pq}(\infty)={1\over 2m\lambda(\lambda^2+\omega^2)}(-\lambda(m\omega)
^2D_{qq}+\lambda D_{pp}+2m\lambda^2D_{pq}).$$
If $\lambda>0$, this evolution law shows an exponential damping of the angular
momentum of the system of three harmonic oscillators, like in [14].

\vskip 0.5truecm

{\bf References}

\item{1.}
E. Davies, Quantum Theory of Open Systems (Academic Press, New York, 1976)

\item{2.}
P. Exner, Open Quantum Systems and Feynman Integrals (Reidel, Dordrecht, 1985)

\item{3.}
R. W. Hasse, J. Math. Phys. {\bf 16} (1975) 2005

\item{4.}
J. Messer, Acta Phys. Austriaca {\bf 58} (1979) 75

\item{5.}
H. Dekker, Phys. Reports {\bf 80} (1981) 1

\item{6.}
K. H. Li, Phys. Reports {\bf 134} (1986) 1

\item{7.}
G. Lindblad, Commun. Math. Phys. {\bf 48} (1976) 119

\item{8.}
G. Lindblad, Rep. on Math. Phys. {\bf 10} (1976) 393

\item{9.}
G. Lindblad, Non-Equilibrium Entropy and Irreversibility (Reidel, Dordrecht,
1983)

\item{10.}
A. Sandulescu, H. Scutaru, Ann. Phys. (N.Y.) {\bf 173} (1987) 277

\item{11.}
A. Sandulescu, H. Scutaru, W. Scheid, J. Phys. A: Math. Gen. {\bf 20} (1987)
2121

\item{12.}
A. Pop, A. Sandulescu, H. Scutaru, W. Greiner, Z. Phys. A - Atomic Nuclei {\bf
329}
(1988) 357

\item{13.}
A. Isar, A. Sandulescu, H. Scutaru, W. Scheid, Nuovo Cimento {\bf 103} (1990)
413

\item{14.}
A. Isar, A. Sandulescu, W. Scheid, Intern. J. Mod. Phys. A{\bf 5} (1990) 1773

\item{15.}
A. Isar, A. Sandulescu, W. Scheid, J. Phys. G: Nucl. Part. Phys. {\bf 17}
(1991) 385

\bye